\documentclass[aps,amsfonts,nofootinbib,superscriptaddress,preprint]{revtex4-1}   

\usepackage{amsmath}
\usepackage{amssymb}
\usepackage{appendix}
\usepackage{bm}
\usepackage{epsfig} 
\usepackage{color}
\usepackage{slashed}

\usepackage[utf8]{inputenc}
\usepackage[T1]{fontenc}
\usepackage{hyperref}
\usepackage{graphicx,bm,amsmath,color,scalerel}
\usepackage{bbold}
\usepackage{wasysym}
\usepackage{verbatim}
\usepackage{amssymb}
\usepackage{epstopdf}
\usepackage{animate}
\usepackage{xmpmulti}
\usepackage{centernot}
\usepackage{multirow}
\usepackage{tikz}
\usepackage{comment}
\usepackage[normalem]{ulem} 
\makeatletter

\newcommand{\be}{\begin{equation}}
\newcommand{\ee}{\end{equation}}
\newcommand{\beq}{\begin{eqnarray}}
\newcommand{\eeq}{\end{eqnarray}}
\newcommand{\ba}{\begin{align}}
\newcommand{\ea}{\end{align}}

\begin{document}

\bigskip


\bigskip
\title{Deformed classical-quantum mechanics transition}

\author{J.L. Cort\'es}
\affiliation{Departamento de F\'{\i}sica Te\'orica and Centro de Astropartículas y Física de Altas Energías (CAPA),
  Universidad de Zaragoza, Zaragoza 50009, Spain}
\author{J. Gamboa}
\affiliation{Departamento de  F\'{\i}sica, Universidad de  Santiago de Chile, Casilla 307, Santiago, Chile}
\begin{abstract}
An approach to study a generalization of the classical-quantum transition for general systems is proposed. In order to develop the idea, a deformation of the ladder operators algebra is proposed that contains a realization of the quantum group $SU(2)_q$ as a particular case. In this deformation Planck's  constant becomes an operator whose eigenvalues approach $\hbar $ for small values of  $n$ (the eigenvalue of the number operator), and zero for large values of $n$ (the system is classicalized).
\end{abstract}

\maketitle

\section{Introduction}

There are different motivations to consider a deformation of the classical-quantum transition \cite{Kempf:1996fz,Kempf:1996nk,Kempf:1994su,Chang:2001bm,AmelinoCamelia:2008qg,Das:2009hs,Capozziello:1999wx,Tawfik:2014zca,Cortes:2004qn,Hossenfelder:2012jw,Ballesteros:2020uxp,Lake:2018zeg}. The difficulties to find a consistent quantum theory incorporating the gravitational interaction could be due to the present formulation of  quantum theories \cite{[See R. Penrose pp.581 in ]Hawking:1979ig}. A modification of this formulation through the classical-quantum transition can provide a new way to overcome those difficulties. Different arguments combining gravity and quantum mechanics \cite{Kibble:1978tm} lead to the conclusion that there is a minimum length \cite{Yoneya:2000bt} and then to consider a non-commutative space \cite{Snyder:1946qz} and then a generalization of the uncertainty principle which implicitly goes beyond the formulation of quantum mechanics \cite{Banks:1983by,Horowitz:1996nw}. There are also arguments that a theory of gravity at super-Planckian energies should become a classical theory \cite{Dvali:2018xoc,Dvali:2014ila,Dvali:2012mx,Hooft:2015ola} which leads to look for a framework able to accommodate such classicality. The difficulties to make compatible the quantum uncertainty with the classical determinism (measurement problem) can also be solved by a modification of the classical-quantum transition. We end up the list of motivations for a deformation of the classical-quantum. transition pointing out the possibility that it provides a new way to try to overcome the difficulties to understand some surprising quantum mechanical effects like the phenomenon of high temperature superconductivity and Bose-Einstein condensation. Indeed standard low temperature superconductivity is successfully explained by the formation of Cooper pairs above a critical temperature $T_C$. But from the point of view of Bose-Einstein condensation there is no way to obtain a critical temperature much higher than $T_C$ unless, a) another Cooper pair formation mechanism takes place, or b) another completely different mechanism, possibly beyond current quantum mechanics occurs. This last possibility is one of the motivations for the present work.

Along the lines outlined in b), much work has been done either modifying the Heisenberg uncertainty relations \cite{Kempf:1996fz,Kempf:1996nk,Kempf:1994su,Chang:2001bm}, re-studying the classical-quantum mechanics transition \cite{Ibort:2016qvv,Asorey:2012cj,Hartle:2015bna,Marmo:2005dc,Mangano:2015pha}, modifying quantum mechanics using quantum groups \cite{Kempf:1993bq} or using arguments from non-commutative geometry \cite{Dunne:1989hv,Gamboa:2000yq,Gamboa:2001qa,Gamboa:2001fg,Falomir:2002ih,Bolonek:2007tk,Kijanka:2004gs,Mezincescu:2000zq,Acatrinei:2001wa,Gomes:2009tk,Bertolami:2009wa,Bastos:2007bg,Bertolami:2005jw,Gomes:2010xk} and related arguments \cite{Kempf:1994su,Wani:2020xus,Park:2020zom,Merad:2019sxl,Bishop:2019yft}.

A strategy to study classical-quantum transition for general systems is proposed. In order to develop the idea, a deformation of the ladder operators algebra is proposed and contains a limit to $ SU(2)_q $ as a symmetry group. In this deformation the Planck constant becomes an operator whose eigenvalues approach $\hbar$ for small values of the quantum number $n$, but for large values of $ n$, the eigenvalues approach zero and the system is classicalized.
A deformation of the classical-quantum transition can be interpreted as an step to go beyond standard quantum mechanics considered (as any other physical theory) as an effective theory which breaks down in some extreme conditions. In this work we try to implement this idea by replacing the fundamental constant in quantum mechanics, the Planck constant $\hbar$, by an operator. In the limit when the operator can be approached by a constant one recovers standard quantum mechanics. One way to replace the Planck constant by an operator, the one we consider in this work, is based on a deformation of the algebra of ladder operators which reproduce the Heisenberg algebra. 

\medskip

In section \ref{DLO} we introduce a deformed algebra of ladder operators in terms of a continuous family of real functions of one variable depending on one continuous deformation parameter ($\lambda$). The corresponding deformed Fock space and a simple example for the deformed algebra are identified. In section \ref{DQM} it is shown how, the formulation of a classical mechanical system in terms of appropriately chosen combinations of the phase space coordinates in complex variables, defines a deformation of the classical-quantum transition through the replacement of the complex variables by deformed ladder operators. The consequences of the deformation are studied in three simple cases: a one and two dimensional harmonic oscillator and a particle in two dimensions in the presence of a constant magnetic field. The deformation of the Heisenberg algebra and the corresponding generalized uncertainty principle for the three cases are discussed in section \ref{DHA} and in section \ref{DUP}. The relation of the proposal presented in this work with previous works where the idea of a deformation of standard quantum mechanics has been considered from different perspectives is discussed in section \ref{COA}. We end up with a summary and future prospects in section \ref{SO}.       

\section{Deformed ladder operators}
\label{DLO}
We start considering deformed creation ($\tilde{a}^\dagger$) and annihilation ($\tilde{a}$) operators such that
\be
[\tilde{a}, \tilde{a}^\dagger] \,=\, {\cal D}_\lambda(\tilde{a}^\dagger \tilde{a}), 
\label{eq:dlo}
\ee
where $\lambda$ is a real number and ${\cal D}_\lambda(x)$ a real function of a real variable ($x$) satisfying the conditions ${\cal D}_0(x)=1$ and ${\cal D}_\lambda(x)>0$. The first condition is what defines a deformation since when $\lambda \to 0$ one recovers the standard creation-annihilation commutation relations, with $a^\dagger$ ($a$) increasing (decreasing) the eigenvalue of $a^\dagger a$ in one unit. The second condition guarantees that $\tilde{a}$, $\tilde{a}^\dagger$ are ladder operators with $\tilde{a}$ ($\tilde{a}^\dagger$) decreasing (increasing) the eigenvalue of $\tilde{a}^\dagger \tilde{a}$.
We also add the condition ${\cal D}_\lambda'(x) < 0$ to have a simple deformation of the spectrum of the operator $\tilde{a}^\dagger \tilde{a}$.

\subsection{Deformed Fock space}
The spectrum of the operator $\tilde{a}^\dagger \tilde{a}$ can be derived from the commutator Eq.~\eqref{eq:dlo}. There is an eigenvalue $c_n(\lambda)$ for each integer $n$ such that  
\be
c_{n+1}(\lambda) \,=\, c_n(\lambda) + {\cal D}_\lambda(c_n(\lambda)).
\label{eq:rr}
\ee
The recurrence relation Eq.~\eqref{eq:rr} defines the eigenvalues if one assumes there is one state ($|0>$) such that $\tilde{a} |0> =0$ corresponding to the lowest eigenvalue ($c_0(\lambda)=0$) of $\tilde{a}^\dagger \tilde{a}$. One has $c_1(\lambda)=1$, $c_{n+1}(\lambda) - c_n(\lambda) < 1$ for $n>0$, and each choice of the function ${\cal D}_\lambda$ leads to a different $\lambda$-dependent contraction of the natural numbers. Alternatively, a different choice of the function defining the deformation such that ${\cal D}_\lambda'(x) > 0$ would have lead to a $\lambda$-dependent expansion of the natural numbers. 

One has a linear representation of the algebra Eq.~\eqref{eq:dlo} in a space (deformed Fock space) with an orthonormal basis that together with the state $|0>$ has the states 
\be
|n> \,=\, [\prod_{j=1}^n c_j(\lambda)]^{-1/2} \, (\tilde{a}^\dagger)^n |0>, 
\label{dFs-basis}
\ee
for any natural number $n$. 

\subsection{Linear deformation}
The simplest example for the function ${\cal D}$ defining the deformed  commutators of ladder operators is a linear function ${\cal D}_\lambda(x)=1-\lambda x$. In this case the recurrence relations Eq.~\eqref{eq:rr} reduce to  
\be
c_{n+1}(\lambda) \,=\, 1 + (1-\lambda) c_n(\lambda), 
\ee
which combined with 
\be
c_n(\lambda) \,=\, 1 + (1-\lambda) c_{n-1}(\lambda)
\ee
leads to 
\be
c_{n+1}(\lambda) - c_n(\lambda) \,=\, (1-\lambda) (c_n(\lambda) - c_{n-1}(\lambda)) \,=\, ...\,=\, (1-\lambda)^n (c_1(\lambda) - c_0(\lambda)) \,=\, (1-\lambda)^n,
\ee
and 
\be
c_n(\lambda) \,=\, \frac{1 - (1-\lambda)^n}{\lambda}.
\ee
In the case $\lambda>0$ one has $c_n(\lambda) < 1/\lambda$, $\lim_{n\to \infty} c_n(\lambda) = 1/\lambda$ and $\lim_{n\to \infty} (c_{n+1}(\lambda) - c_n(\lambda)) = 0$. One has a bounded spectrum for the operator $\tilde{a}^\dagger \tilde{a}$ with the inverse of the deformation parameter playing the role of a cutoff and the discrete spectrum approaches a continuum spectrum for large $n$ (classicality). 

The deformed commutation relations of ladder operators Eq.~\eqref{eq:dlo} can be written in this case as
\be
\tilde{a} \tilde{a}^\dagger - (1-\lambda) \tilde{a}^\dagger \tilde{a} \,=\, 1, 
\ee
which is known as a q-commutator~\cite{Biedenharn:1989jw,Macfarlane:1989dt,Chaichian:1990up} with the identification $q^2 = 1-\lambda$. The case $q^2 > 1$ ($\lambda < 0$) is the one-dimensional case of the studies of deformations of quantum mechanics with a quantum group symmetry~\cite{Kempf:1994su}. In this work we will be more interested in the case $\lambda>0$.  

\section{Deformed Quantum Mechanics}
\label{DQM}

In order to identify a deformation of the classical-quantum transition from the deformed ladder operator commutation relations, we need to reformulate a classical system in terms of complex variables $\alpha_i$ (one for each degree of freedom, $i=1, ...,n$) which will become the ladder operators in the quantum theory (holomorphic representation) \cite{Perelomov:1986tf}. This requires to identify linear combinations of the phase space variables such that the quadratic part of the classical hamiltonian $h^{(2)}$ takes the form
\be
h^{(2)} \,=\, \sum_{i=1}^n \, \epsilon_i \, \alpha^*_i \alpha_i.
\ee
In the case without deformation this is just the hamiltonian of n harmonic oscillators with frequencies $\omega_i$ when $\epsilon_i = \hbar \omega_i$. If one has a particle in an external potential with a non-degenerate minimum, then the holomorphic representation can easily be obtained from the diagonalization of the matrix whose elements are the second derivatives of the potential at the minimum. 

The case of a harmonic oscillator is just the particular case where the potential is quadratic in the space coordinates. In more general cases one has to assume a Taylor expansion of the potential around its minimum and the hamiltonian will have, together with the quadratic terms ($h^{(2)}$), higher powers of the space coordinates which can be expressed as products of the complex variables $\alpha_i$. In next section we will show a few simple examples of holographic representations.   

Once the hamiltonian of the classical system has been written in terms of complex variables one can define the hamiltonian of the quantum system ($H$) as the operator obtained by replacing the variables $\alpha_i$ by operators $\tilde{a}_i$ satisfying the algebra
\be
[\tilde{a}_i, \tilde{a}^\dagger_j] \,=\, \delta_{ij} \, {\cal D}_\lambda(\tilde{a}^\dagger_i \tilde{a}_i), \quad\quad\quad [\tilde{a}_i, \tilde{a}_j] \,=\, [\tilde{a}^\dagger_i, \tilde{a}^\dagger_j] \,=\, 0.
\label{def-algebra}
\ee
If one wants to go beyond the linear choice for the function ${\cal D}$ one has to restrict to a decoupled algebra for each degree of freedom~\footnote{In the case of a linear algebra (with $\lambda < 0$) it is possible to consider a deformed algebra (quantum group $SU_q(n)$ algebra) with a mixing of different degrees of freedom.}. There is an ordering ambiguity which can be fixed writing all factors $\alpha^*_i$ to the left of factors $\alpha_i$ in the classical hamiltonian $h$ so that one has a normal ordered quantum hamiltonian $H$.     

\subsection{One-dimensional harmonic oscillator}
The hamiltonian of the classical system is in this case
\be
h \,=\, \frac{p^2}{2 m} + \frac{m \omega^2 x^2}{2} \,=\, \hbar \omega \alpha^* \alpha,
\ee
with 
\be
\alpha \,=\, \frac{1}{\sqrt{2}} \,\left({\sqrt\frac{m \omega}{\hbar}} \: x + i \, {\sqrt \frac{1}{\hbar m \omega}} \: p\right).
\label{D=1,a}
\ee
If one replaces the variable $\alpha$ by an operator $\tilde{a}$ which together with $\tilde{a}^\dagger$ satisfy the deformed algebra Eq.~\eqref{eq:dlo} then the (normal ordered) deformed quantum oscillator hamiltonian has the spectrum
\be
E_n \,=\, \hbar \omega \, c_n(\lambda),
\ee
instead of the equally spaced standard quantum oscillator spectrum. The contraction (expansion) of the natural numbers defined by $c_n(\lambda)$ when $\lambda>0$ ($\lambda<0$) leads to a contraction (expansion) of the spectrum of the deformed quantum oscillator. In the case of a linear deformation one has a splitting of energy levels
\be
E_{n+1} - E_n \,=\, \hbar \omega (1 - \lambda)^n
\ee
which approaches to a continuum spectrum (classicality) at large $n$ when $\lambda>0$. 

The standard construction of coherent states in Fock space can be generalized to coherent deformed states as the eigenvectors $|\alpha>$ of the operator $\tilde{a}$, $\tilde{a} |\alpha> = \alpha |\alpha>$. One has
\be
|\alpha> \,=\, {\cal N}(\alpha) \left[\,|0> + \sum_{n=1}^\infty \, \frac{\alpha^n}{\left[\prod_{j=1}^n c_j(\lambda)\right]^{1/2}} \, |n>\right].
\ee
and one has $<\alpha|\alpha>=1$ if one chooses 
\be
{\cal N}(\alpha) \,=\, \left({\cal D}\exp{[|\alpha|^2}]\right)^{-1/2},
\ee
where 
\be
{\cal D}\exp{[x]} \,\doteq\, 1 + \sum_{n=1}^\infty \,\frac{x^n}{\left[\prod_{j=1}^n c_j(\lambda)\right]}
\ee
is the deformed exponential.


\subsection{Two-dimensional harmonic oscillator}
We consider a two-dimensional isotropic oscillator with a classical hamiltonian
\be
h \,=\, \frac{p_1^2}{2 m} + \frac{p_2^2}{2 m} + \frac{m \omega^2 x_1^2}{2} + \frac{m \omega^2 x_2^2}{2} \,=\, \hbar \omega \,\left[\alpha_1^* \alpha_1 + \alpha_2^* \alpha_2\right].
\ee
A new ingredient with respect to the one-dimensional case is that there is some ambiguity in the identification of the complex variables $\alpha_1$, $\alpha_2$ as linear combinations of the phase space coordinates. Each choice of these variables leads to a different quantum system. 

The simplest choice 
\be
\alpha_1 \,=\, \frac{1}{\sqrt{2}} \,\left({\sqrt\frac{m \omega}{\hbar}} \: x_1 + i \, {\sqrt \frac{1}{\hbar m \omega}} \: p_1\right), \quad\quad\quad
\alpha_2 \,=\, \frac{1}{\sqrt{2}} \,\left({\sqrt\frac{m \omega}{\hbar}} \: x_2 + i \, {\sqrt \frac{1}{\hbar m \omega}} \: p_2\right),
\ee
leads to a quantum system where the rotation with angle $\theta$ 
\begin{align}
x'_1 \,=\, \cos\theta \, x_1 + \sin\theta \, x_2, \quad\quad\quad
x'_2 \,=\, -\sin\theta \, x_1 + \cos\theta \, x_2, \nonumber \\
p'_1 \,=\, \cos\theta \, p_1 + \sin\theta \, p_2, \quad\quad\quad
p'_2 \,=\, -\sin\theta \, p_1 + \cos\theta \, p_2,
\end{align}
acts on the complex variables as
\be
\alpha'_1 \,=\, \cos\theta \, \alpha_1 + \sin\theta \, \alpha_2, \quad\quad\quad
\alpha'_2 \,=\, -\sin\theta \, \alpha_1 + \cos\theta \, \alpha_2\,.
\ee
The deformed algebra~(\ref{def-algebra}) is not invariant under the corresponding transformation on the operators $\tilde{a}_1$, $\tilde{a}_2$ and the rotational symmetry of the classical system is lost in the classical-quantum transition. In order to maintain the rotational symmetry in the quantum system one has to choose
\begin{align}
& \alpha_1 \,=\, \frac{1}{2} \,\left({\sqrt\frac{m \omega}{\hbar}} \: (x_1 + i x_2) + i \, {\sqrt \frac{1}{\hbar m \omega}} \: (p_1 + i p_2)\right), \nonumber \\
& \alpha_2 \,=\, \frac{1}{2} \,\left({\sqrt\frac{m \omega}{\hbar}} \: (x_1 - i x_2) + i \, {\sqrt \frac{1}{\hbar m \omega}} \: (p_1 - i p_2)\right).
\end{align}
In this case one has
\be
\alpha'_1 \,=\, e^{-i\theta} \, \alpha_1, \quad\quad\quad   \alpha'_2 \,=\, e^{i\theta} \, \alpha_2\,,
\ee
and the correponding transformation of the deformed operators $\tilde{a}_1$, $\tilde{a}_2$ leaves the algebra~(\ref{def-algebra}) invariant.  
The spectrum of the quantum hamiltonian is 
\be
E_{n_1, n_2} \,=\, \hbar \omega [c_{n_1}(\lambda) + c_{n_2}(\lambda)]\,,
\ee
generalizing the spectrum of the two-dimensional quantum oscillator which is reproduced if one replaces $c_n(\lambda)$ by $n$. 

\subsection{Landau quantization}
The next example we consider of a deformed classical-quantum transition is the classical system of a particle in two-dimensional space in the presence of a constant magnetic field $B$. The classical hamiltonian is
\be
h \,=\, \frac{(p_1 - q A_1)^2}{2 m} \,+\, \frac{(p_2 - q A_2)^2}{2 m}\,, 
\ee
where ($A_1$, $A_2$) is the electromagnetic potential corresponding to the magnetic field, $B = \partial_1 A_2 - \partial_2 A_1$. There are different choices for the electromagnetic potential (different gauges). 

In the symmetric gauge one has $A_1 = - x_2/2$, $A_2 = x_1/2$ and then 
\be
h \,=\, \frac{\left(p_1 + \frac{q B}{2} x_2\right)^2}{2 m} \,+\, \frac{\left(p_2 - \frac{q B}{2} x_1\right)^2}{2 m} \,=\, \hbar \omega \alpha^* \alpha\,,
\ee
where $\omega = (q B)/m$ and 
\be
\alpha \,=\, \frac{1}{2} \,\left({\sqrt\frac{m \omega}{2 \hbar}} \: (x_1 + i x_2) + i \, {\sqrt \frac{2}{\hbar m \omega}} \: (p_1 + i p_2)\right)\,.
\label{a-symmetric}
\ee
It is convenient to introduce a second complex linear combinations of the phase space variables $\beta$, orthogonal to $\alpha$, 
\be
\beta \,=\, \frac{1}{2} \,\left({\sqrt\frac{m \omega}{2 \hbar}} \: (x_1 - i x_2) + i \, {\sqrt \frac{2}{\hbar m \omega}} \: (p_1 - i p_2)\right)\,.
\ee
The deformed classical quantum transition is defined by replacing the complex variables $\alpha$, $\alpha^*$ in the hamiltonian by operators $\tilde{a}$, $\tilde{a}^\dagger$ satisfying the deformed algebra Eq.~\eqref{eq:dlo}. The spectrum of the deformed quantum hamiltonian will be 
\be
E_n \,=\, \hbar \omega \,c_n(\lambda) \,=\, \hbar \frac{q B}{m} \, c_n(\lambda)
\ee
The ladder operators $\tilde{a}$, $\tilde{b}$ are similar to the operators $\tilde{a}_1$, $\tilde{a}_2$ found in the case of the two-dimensional oscillator in order to maintain the rotational symmetry in the classical-quantum transition. The difference is that in the hamiltonian there is only one term instead of two terms with the same frequency.

In the Landau gauge one has $A_1 = - B x_2$, $A_2 = 0$ and the hamiltonian is given by 
\be
h \,=\, \frac{\left(p_1 + q B x_2\right)^2}{2 m} \,+\, \frac{p_2^2}{2 m} \,=\, \frac{p_2^2}{2 m} \,+\, \frac{1}{2} m \left(\frac{q B}{m}\right) ^2 \left(x_2 + \frac{p_1}{q B}\right)^2 \,=\, \hbar \omega \,\alpha^* \alpha,
\ee
with $\omega = (q B)/m$. The complex variable $\alpha$ is in this case
\be 
\alpha \,=\, \frac{1}{\sqrt{2}} \,\left(\sqrt{\frac{m\omega}{\hbar}} \,x_2 + \sqrt{\frac{1}{\hbar m \omega}}\, (p_1 + i p_2)\right)
\ee
instead of (\ref{a-symmetric}). The orthogonal complex linear combination of phase space variables $\beta$ is in this case
\be
\beta \,=\, \frac{1}{\sqrt{2}} \,\left(\sqrt{\frac{m\omega}{\hbar}} \,x_1 + \sqrt{\frac{1}{\hbar m \omega}}\, (p_2 + i p_1)\right).
\ee

The spectrum of the quantum hamiltonian is the same in both gauges with
\be
E_n \,=\, \hbar \frac{q B}{m} \, c_n(\lambda).
\ee


\subsection{Rotational symmetry}
In the two examples of two-dimensional quantum systems (harmonic oscillator and a particle in a constant magnetic field) we have found that it is possible to make the deformation compatible with the rotational symmetry of the quantum system. In the first case this is done using the arbitrariness in the holomorphic representation of the classical system and in the second case using the appropriate choice of gauge (symmetric gauge). 

If one considers the three dimensional harmonic oscillator it is not possible to make compatible the deformation in the transition from the classical to the quantum system with rotational symmetry. All one could do is to choose the holomorphic representation such that one has a symmetry under rotations in a given direction in the quantum system. If one considers a very small deformation parameter ($\lambda \ll 1$), the effect of the deformation (and then the violation of rotational symmetry) in the spectrum starts to be appreciable when $n$ is sufficiently large that $c_n(\lambda)$ differs from $n$. On the other hand, if one goes to still much larger values of $n$, one will approach the classical continuum limit.

\section{Deformed Heisenberg algebra}
\label{DHA}

Let us start with the one-dimensional deformed quantum harmonic oscillator. The commutator of operators which define the Heisenberg algebra is given by
\be
[x, p] \,=\, \left[\sqrt{\frac{\hbar}{2 m \omega}} (\tilde{a} + \tilde{a}^\dagger), - i \sqrt{\frac{\hbar m \omega}{2}} (\tilde{a} - \tilde{a}^\dagger)\right] \,=\,  i \hbar \left[\tilde{a}, \tilde{a}^\dagger\right] \,=\, i \hbar \,{\cal D}_\lambda(\tilde{a}^\dagger \tilde{a})\,.
\ee
If one uses the basis of eigenstates ($|n>$) of the quantum hamiltonian one has
\be
<n| [x, p] |m> \,=\, \delta_{n, m} \, i \hbar \, {\cal D}_\lambda(c_n(\lambda))\,.
\ee
The effect of the deformation in the Heisenberg algebra is to replace the Planck constant $\hbar$ by an effective (energy dependent) Planck constant
\be
\hbar_{eff}(n) \,\doteq\, \hbar \, {\cal D}_\lambda(c_n(\lambda)) \,=\, \hbar \, \left[c_{n+1}(\lambda) - c_n(\lambda)\right]\,.
\ee
In the case of a linear deformation one has $\hbar_{eff}(n) = \hbar (1-\lambda)^n$ so that when $\lambda >0$ one has $\lim_{n\to \infty} \hbar_{eff}(n) = 0$ (classicality).  

Next we can consider the deformation of the two-dimensional harmonic oscillator compatible with the rotational symmetry. In this case one has the operators corresponding to the phase space variables 
\begin{align}
x_1 \,=\, \frac{1}{2} \, \sqrt{\frac{\hbar}{m \omega}} \,\left(\tilde{a}_1 + \tilde{a}_2 + \tilde{a}_1^\dagger + \tilde{a}_2^\dagger\right), \quad\quad\quad
x_2 \,=\, - \frac{i}{2} \, \sqrt{\frac{\hbar}{m \omega}} \,\left(\tilde{a}_1 - \tilde{a}_2 - \tilde{a}_1^\dagger + \tilde{a}_2^\dagger\right), \nonumber \\
p_1 \,=\, - \frac{i}{2} \, \sqrt{\hbar m \omega} \,\left(\tilde{a}_1 + \tilde{a}_2 - \tilde{a}_1^\dagger - \tilde{a}_2^\dagger\right), \quad\quad\quad
p_2 \,=\, \frac{1}{2} \, \sqrt{\hbar m \omega} \, \left(\tilde{a}_2 - \tilde{a}_1 + \tilde{a}_2^\dagger - \tilde{a}_1^\dagger\right), 
\end{align}
and the deformed commutators
\begin{align}
& [x_1, p_1] \,=\, [x_2, p_2] \,=\, \frac{i \hbar}{2} \,\left([\tilde{a}_1, \tilde{a}_1^\dagger] + [\tilde{a}_2, \tilde{a}_2^\dagger]\right) \,=\, \frac{i \hbar}{2} \,\left({\cal D}_\lambda(\tilde{a}_1^\dagger \tilde{a}_1) + {\cal D}_\lambda(\tilde{a}_2^\dagger \tilde{a}_2)\right), \quad [x_1, p_2] \,=\, [x_2, p_1] \,=\, 0, \nonumber \\ 
& [x_1, x_2] \,=\, \frac{i \hbar}{2 m \omega} \,\left([\tilde{a}_1, \tilde{a}_1^\dagger] - [\tilde{a}_2, \tilde{a}_2^\dagger]\right) \,=\, \frac{i \hbar}{2 m \omega} \,\left({\cal D}_\lambda(\tilde{a}_1^\dagger \tilde{a}_1) - {\cal D}_\lambda(\tilde{a}_2^\dagger \tilde{a}_2)\right), \nonumber \\  
& [p_1, p_2] \,=\, \frac{i \hbar m \omega}{2} \,\left([\tilde{a}_1, \tilde{a}_1^\dagger] - [\tilde{a}_2, \tilde{a}_2^\dagger]\right) \,=\, \frac{i \hbar m \omega}{2} \,\left({\cal D}_\lambda(\tilde{a}_1^\dagger \tilde{a}_1) - {\cal D}_\lambda(\tilde{a}_2^\dagger \tilde{a}_2)\right).
\end{align}
If we use the eigenstates ($|n_1, n_2>$) of the quantum two dimensional harmonic oscillator hamiltonian we have
\begin{align}
& <n_1, n_2| [x_1, p_1] |m_1, m_2> \,=\, <n_1, n_2| [x_2, p_2] |m_1, m_2> \,=\, \delta_{n_1, m_1} \delta_{n_2, m_2} \,\frac{i \hbar}{2} \,\left[{\cal D}_\lambda(c_{n_1}(\lambda)) + {\cal D}_\lambda(c_{n_2}(\lambda))\right], \nonumber \\
& <n_1, n_2| [x_1, x_2] |m_1, m_2> \,=\, \delta_{n_1, m_1} \delta_{n_2, m_2} \,\frac{i \hbar}{2 m \omega} \,\left[{\cal D}_\lambda(c_{n_1}(\lambda)) - {\cal D}_\lambda(c_{n_2}(\lambda))\right], \nonumber \\
& <n_1, n_2| [p_1, p_2] |m_1, m_2> \,=\, \delta_{n_1, m_1} \delta_{n_2, m_2} \,\frac{i \hbar m \omega}{2} \,\left[{\cal D}_\lambda(c_{n_1}(\lambda)) - {\cal D}_\lambda (c_{n_2}(\lambda))\right].
\end{align}
Together with an effective (energy dependent) Planck constant
\be
\hbar_{eff}(n_1, n_2) \,\doteq\, \frac{\hbar}{2} \,\left[{\cal D}_\lambda(c_{n_1}(\lambda)) + {\cal D}_\lambda(c_{n_2}(\lambda))\right] \,=\, \frac{\hbar}{2} \,\left[c_{n_1+1}(\lambda) - c_{n_1}(\lambda) + c_{n_2+1}(\lambda) - c_{n_2}(\lambda)\right],
\ee
one has, as a consequence of the deformation in the transition from the classical to the quantum system, a non-commutativity of space operators and also of momentum operators with (energy dependent) space (momentum) non-commutativity parameters $\theta(n_1, n_2)$ (${\cal B}(n_1, n_2)$) 
\begin{align}
\theta(n_1, n_2) \,\doteq\, \frac{\hbar}{2 m \omega} \,\left[c_{n_1+1}(\lambda) - c_{n_1}(\lambda) - c_{n_2+1}(\lambda) + c_{n_2}(\lambda)\right], \nonumber \\
{\cal B}(n_1, n_2) \,\doteq\, \frac{\hbar m \omega}{2} \,\left[c_{n_1+1}(\lambda) - c_{n_1}(\lambda) - c_{n_2+1}(\lambda) + c_{n_2}(\lambda)\right].
\end{align}
Once more one can see that when $\lambda > 0$ one has a classicality limit \be
\lim_{n_1, n_2 \to \infty} \hbar_{eff}(n_1, n_2) \,=\,  \lim_{n_1, n_2 \to \infty} \theta_{eff}(n_1, n_2) \,=\, \lim_{n_1, n_2 \to \infty} {\cal B}_{eff}(n_1, n_2) \,=\, 0.
\ee

All the results of the deformed Heisenberg algebra for the two-dimensional harmonic oscillator can be applied to the case of a particle in two dimensions in the presence of a constant magnetic field when one uses the symmetric gauge. All one has to do is to replace everywhere the frequency $\omega$ of the harmonic oscillator by the ratio $(q B/m)$. The space (momentum) non-commutativity parameters satisfy in this case the relations
\be
\frac{q B}{\hbar} \, \theta(n_1, n_2) \,=\,  \frac{1}{\hbar q B} \, {\cal B}(n_1, n_2) \,=\, \frac{1}{2} \,\left[c_{n_1+1}(\lambda) - c_{n_1}(\lambda) - c_{n_2+1}(\lambda) + c_{n_2}(\lambda)\right],
\ee
where $n_1$ is the integer which fixes the energy levels and $n_2$ is the integer which specifies the different states in each energy level.  

\section{Deformed uncertainty principle}
\label{DUP}

From the expression for the commutator $[x, p]$ in the one dimensional case we conclude that for any state $|\Psi>$ one has the lower bound for the product of the uncertainties in the position and momentum 
\be
(\Delta x)_\Psi \, (\Delta p)_\Psi \geq \frac{\hbar}{2} \,|<\Psi| {\cal D}_\lambda( \tilde{a}^\dagger \tilde{a}) |\Psi>|.
\ee
If one considers the eigenstates $|n>$ of the product of ladder operators $\tilde{a}^\dagger \tilde{a}$ one has
\be
(\Delta x)_n \, (\Delta p)_n \geq \frac{\hbar}{2} \left[c_{n+1}(\lambda) - c_n(\lambda)\right]\,.
\ee
When $\lambda >0$ the lower bound decreases when $n$ increases and it can be made arbitrarily small. This is an indication that there will be states where one can make the uncertainties in the position and momentum operators arbitrarily small. If we calculate directly the uncertainty of the position operator in the state $|n>$ we have
\begin{align}
(\Delta x)_n^2 &=\, <n| x^2 |n> - <n| x |n>^2 \,=\, \frac{\hbar}{2 m \omega} \,<n| (\tilde{a} + \tilde{a}^\dagger)^2 |n> \,=\, \frac{\hbar}{2 m \omega} \,<n| (2 \tilde{a}^\dagger \tilde{a} + [\tilde{a}, \tilde{a}^\dagger])|n> \nonumber \\ &=\, \frac{\hbar}{2 m \omega} \,\left[2 c_n(\lambda) + {\cal D}_\lambda(c_n(\lambda))\right] \,=\, \frac{\hbar}{2 m \omega} [c_n(\lambda) + c_{n+1}(\lambda)]\,.
\end{align}
Then these are not the states we are looking for. If we consider the coherent states $|\alpha>$ we have 
\be
(\Delta x)_\alpha^2 \,=\, \frac{\hbar}{2 m \omega} \,\left[<\alpha|(\tilde{a} + \tilde{a}^\dagger)^2|\alpha> - <\alpha|(\tilde{a} + \tilde{a}^\dagger)|\alpha>^2\right] \,=\, \frac{\hbar}{2 m \omega} \,<\alpha|[\tilde{a}, \tilde{a}^\dagger]|\alpha>. 
\ee
In the case of the linear deformation, the uncertainty of the position operator in a coherent state is
\be
(\Delta x)_\alpha^2 \,=\, \frac{\hbar}{2 m \omega} \,\left(1 - \lambda |\alpha|^2\right) 
\ee
and one has coherent states for any complex number $\alpha$ such that $0<|\alpha|^2<1/\lambda$. Then one has 
\be
\lim_{|\alpha|^2 \to 1/\lambda} \,(\Delta x)_\alpha^2 \,=\, 0.
\ee
A similar analysis can be made for the uncertainty of the momentum operator in a coherent state. The result is
\be
(\Delta p)_\alpha^2 \,=\, \frac{\hbar m \omega}{2} \,\left(1 - \lambda |\alpha|^2\right),
\ee
and one also has 
\be
\lim_{|\alpha|^2 \to 1/\lambda} \,(\Delta p)_\alpha^2 \,=\, 0.
\ee
Then one has states where both the uncertainties in the position and momentum operator can be made arbitrarily small (classicality). One would expect this will have implications on the issue of locality in quantum mechanics \cite{Einstein:1935rr,Hardy:1993zza}.

\smallskip

In the two-dimensional harmonic oscillator one can also consider the uncertainty of a position operator in a two-dimensional coherent state $|\alpha_1, \alpha_2>$ 
\begin{align}
(\Delta x_1)^2_{\alpha_1,\alpha_2} &=\, \frac{\hbar}{4 m \omega} \,\left[<\alpha_1,\alpha_2| \left(\tilde{a}_1 + \tilde{a}_2 + \tilde{a}_1^\dagger + \tilde{a}_2^\dagger\right)^2 |\alpha_1,\alpha_2> - <\alpha_1,\alpha_2| \left(\tilde{a}_1 + \tilde{a}_2 + \tilde{a}_1^\dagger + \tilde{a}_2^\dagger\right) |\alpha_1,\alpha_2>^2\right] \nonumber \\
 &=\, <\alpha_1,\alpha_2| [\tilde{a}_1, \tilde{a}_1^\dagger] |\alpha_1,\alpha_2> + <\alpha_1,\alpha_2| [\tilde{a}_2, \tilde{a}_2^\dagger] |\alpha_1,\alpha_2> \,.
\end{align}
In the case of a linear deformation one has
\be
(\Delta x_1)^2_{\alpha_1,\alpha_2} \,=\, \frac{\hbar}{4 m \omega} \,\left[(1 - \lambda |\alpha_1|^2) + (1 - \lambda |\alpha_2|^2)\right],
\ee
and 
\be
\lim_{|\alpha_1|^2\to 1/\lambda} \:\: \lim_{|\alpha_2|^2\to 1/\lambda} \,(\Delta x_1)^2_{\alpha_1,\alpha_2} \,=\, 0.
\ee
The same result applies for $(\Delta x_2)^2_{\alpha_1,\alpha_2}$. 

For the uncertainties of momentum operators one has
\be
(\Delta p_1)^2_{\alpha_1,\alpha_2} \,=\, (\Delta p_2)^2_{\alpha_1,\alpha_2} \,=\, \frac{\hbar m \omega}{4} \,\left[(1 - \lambda |\alpha_2|^2) - (1 - \lambda |\alpha_1|^2)\right]
\ee
and then 
\be
\lim_{(|\alpha_1|^2\to 1/\lambda} \:\: \lim_{|\alpha_2|^2\to 1/\lambda} \,(\Delta p_1)^2_{\alpha_1,\alpha_2} \,=\, \lim_{(|\alpha_1|^2\to 1/\lambda} \:\: \lim_{|\alpha_2|^2\to 1/\lambda} \,(\Delta p_2)^2_{\alpha_1,\alpha_2} \,=\,0.
\ee

Finally let us mention that all the results of the two-dimensional harmonic oscillator apply to the Landau system with the identification $m \omega = q B$. 

\section{Comparison with other approaches to the deformation of the classical-quantum transition}
\label{COA}

In order to study the relation of the proposal to consider a deformation of the classical-quantum transition based on deformed ladder operators satisfying the algebra~\eqref{eq:dlo} with previous works on deformed quantum mechanics~\cite{Biedenharn:1989jw,Macfarlane:1989dt,Chaichian:1990up}, it is convenient to introduce an operator $N$ such that
\be
[N, \tilde{a}] \,=\, - \tilde{a}\,, \quad\quad\quad
[N, \tilde{a}^\dagger] \,=\, \tilde{a}^\dagger\,,
\ee
so that its eigenvalues differ by integer numbers. Then one can introduce new operators $a$, $a^\dagger$ such that 
\be
N \,=\, a^\dagger a\,, \quad\quad\quad [a, a^\dagger] \,=\, 1\,\quad\quad\quad [N, a] \,=\, -a\,, \quad\quad\quad [N, a^\dagger] \,=\, a^\dagger\,.
\ee
The comparison of the commutators of the operator $N$ with ($\tilde{a}$, $\tilde{a}^\dagger$) and with ($a$, $a^\dagger$) leads to the nonlinear relations 
\be
\tilde{a} \,=\, a f(N)\,, \quad\quad\quad \tilde{a}^\dagger \,=\, f(N) a^\dagger\,,
\label{f}
\ee
which define deformed ladder operators for each choice of the function f.

\medskip

One has 
\be
\tilde{a}^\dagger \tilde{a} \,=\, f(N) a^\dagger a f(N) \,=\, N f^2(N)\,,
\ee
and the eigenvalues ($c_n(\lambda)$) of the operator $\tilde{a}^\dagger \tilde{a}$ are just the eigenvalues of $N f^2(N)$, i.e., $n f^2(n)$. Then one has a correspondence between the spectrum of the deformed quantum oscillator and the function $f$ which defines the nonlinear transformation (\ref{f}) between the ladder operators $a$, $a^\dagger$ and the deformed ladder operators $\tilde{a}$, $\tilde{a}^\dagger$.

\medskip

The deformed algebra of ladder operators in~\eqref{eq:dlo} can then be restated in a more contrived way as
\be
[N, \tilde{a}] \,=\, - \tilde{a}\,, \quad\quad\quad 
[N, \tilde{a}^\dagger] \,=\, \tilde{a}^\dagger\,, \quad\quad\quad [\tilde{a}, \tilde{a}^\dagger] \,=\, F(N)\,,
\label{eq:q-oscillator}
\ee
with the identification 
\be
F(N) \,=\, {\cal D}_\lambda(N f^2(N))\,,
\ee
and the simple recurrence relations~\eqref{eq:rr} for the eigenvalues of the operator $\tilde{a}^\dagger \tilde{a}$ become 
\be
(n+1) f^2(n+1) \,=\, n f^2(n) + F(n)\,.
\ee

A particular choice for the spectrum of the operator $\tilde{a}^\dagger \tilde{a}$ is
\be
c_n(\lambda) \,=\, n f^2(n) \,=\, \frac{e^{\lambda n} - e^{-\lambda n}}{e^\lambda - e^{-\lambda}}\,.
\ee
In this case one has 
\be
F(N) \,=\, (N+1) f^2(N+1) - N f^2(N) \,=\, \frac{\left(e^\lambda -1\right) e^{\lambda N} + \left(1 - e^{-\lambda}\right) e^{-\lambda N}}{e^\lambda - e^{-\lambda}}\,.
\ee
One can also calculate 
\be
\tilde{a} \tilde{a}^\dagger - e^\lambda \tilde{a}^\dagger \tilde{a} \,=\, (N+1) f^2(N+1) - e^\lambda N f^2(N) \,=\, e^{-\lambda N}
\ee
which is the algebraic relation which defines what is known as q-oscillator (with the identification $q=e^\lambda$), used to generalize the representation of the $SU(2)$ group in terms of the ladder operators of two oscillators to the case of the quantum group $SU(2)_q$.

One can also use the relation 
\be
\tilde{a}^\dagger \tilde{a} = \frac{e^{\lambda N} - e^{-\lambda N}}{e^\lambda - e^{-\lambda}}
\ee
to express the operator $N$ in terms of $\tilde{a}^\dagger \tilde{a}$ and then, to identify the function ${\cal D}_\lambda$ which defines (through \eqref{eq:dlo}) the deformed ladder operators corresponding to the q-oscillator
\be
{\cal D}_\lambda(\tilde{a}^\dagger \tilde{a}) \,=\, F(N) \,=\, \left[\sinh^2\lambda (\tilde{a}^\dagger \tilde{a})^2 + 1\right]^{1/2} + (\cosh\lambda -1) \tilde{a}^\dagger \tilde{a}\,.
\ee
But this function is such that ${\cal D}'_\lambda(x) >0$ and then one has that the separation of eigenvalues $c_{n+1}(\lambda) - c_n(\lambda) > 1$, in contrast to the deformation of the classical to quantum transition that approaches a continuum spectrum in the large $n$ limit.  

Another case where one can easily find the spectrum of the deformed quantum oscillator and the associated nonlinear transformation defining the deformed ladder operators is the case we referred to as linear deformation. In this case one has 
\be
f(N) \,=\, \sqrt{\frac{1-(1-\lambda)^N}{\lambda N}}
\ee
for the function defining the nonlinear transformation defining the deformed ladder operator, and 
\be
F(N) \,=\, (N+1) f^2(N+1) - N f^2(N) \,=\, (1-\lambda)^N
\ee
for the function which defines the commutator of ladder operators in terms of the operator $N$. 
We think that one has then a clear relation of the linear deformation with quantum groups. Whether one can identify a generalization of these group structures related with other deformations  is beyond our present knowledge.

The deformation of the classical to quantum transition in the case of a system with two degrees of freedom (either when one has a two-dimensional harmonic oscillator or a particle in the presence of a constant magnetic field) can not be put in correspondence with the  realization of the quantum $SU(2)_q$ quantum group in terms of two q-oscillators. In the deformation of the classical to quantum transition the two pairs of deformed ladder operators ($\tilde{a}_i$, $\tilde{a}_i^\dagger$) do not mix in the algebraic relations~\footnote{See (\ref{def-algebra})} while the algebra of the two q-oscillators which realize the $SU(2)_q$ quantum group requires a mixing. Both operators $N_1$, $N_2$, defined by the conditions
\be
[N_i, \tilde{a}_i] \,=\, - \tilde{a}_i\, \quad\quad\quad 
[N_i, \tilde{a}^\dagger_i] \,=\, \tilde{a}^\dagger_i\,,
\ee
have to appear in the commutators
\be
[\tilde{a}_i, \tilde{a}^\dagger_j] \,=\, \delta_{ij} F(N_1-N_2)\,.
\ee
Then the deformed quantum systems that we have proposed through the deformation of the classical to quantum transition differs from the deformed quantum system associated to the realizations of quantum groups. 

Another related discussion of a possible deformation of a quantum system is based on considering a quantum group $SU(n)_q$ symmetric Fock space defined by the algebraic relations \cite{Kempf:1993bq,Kempf:1992re}

\beq
\tilde{a}_i \tilde{a}_j - q \tilde{a}_j \tilde{a}_i \,=\, 0 \quad (i<j) &&\quad\quad\quad \tilde{a}_i^\dagger \tilde{a}_j^\dagger - q \tilde{a}_j^\dagger \tilde{a}_i^\dagger \,=\, 0 \quad (i>j) \quad\quad\quad \tilde{a}_i \tilde{a}^\dagger_j - q \tilde{a}^\dagger_i \tilde{a}_j \,=\, 0 \quad (i\neq j) \nonumber \\ 
&&\tilde{a}_i \tilde{a}^\dagger_i - q^2 \tilde{a}^\dagger_i \tilde{a}_i \,=\, 1 + (q^2-1) \sum_{j<i} \tilde{a}_j^\dagger \tilde{a}_j
\eeq
which also requires to go beyond the deformation of the classical to quantum transition with algebraic relations which do not mix different pairs of deformed ladder operators. Only if one considers the reduction of the previous algebraic relations to the case of a single pair of operators  
\be
\tilde{a} \tilde{a}^\dagger - q^2 \tilde{a}^\dagger \tilde{a} \,=\, 1
\ee
one can rewrite it as 
\be
[\tilde{a}, \tilde{a}^\dagger] \,=\, 1 - \lambda \tilde{a}^\dagger \tilde{a} 
\ee
with the identification $\lambda = 1 - q^2$. This is just the case referred to as linear deformation with ${\cal D}_\lambda(x) = 1 - \lambda x$ but the quantum group symmetry of the Fock space requires $q^2 > 1$ and then $\lambda <0$ which does not lead to classicality in the large $n$ limit.

We end up pointing out that the spectrum of the deformed quantum oscillator can be identified with the spectrum of a deformed Hamiltonian
\be
\tilde{H} \,=\, H \,f^2(H/\hbar\omega)
\ee
where $H$ is the (normal ordered) quantum Hamiltonian of the harmonic oscillator whose spectrum is given by $E_n = \hbar \omega n$. The spectrum of $\tilde{H}$ is then 
\be
\tilde{E}_n \,=\, \hbar \omega \,n f^2(n) \,=\, \hbar \omega \,c_n(\lambda)
\ee
which is the spectrum of the deformed quantum oscillator. 

\section{Summary and outlook}
\label{SO}

We have presented in this work a proposal for the deformation of the ladder operators associated with the Heisenberg algebra of a quantum mechanical system. In the case of a classical mechanical system with a hamiltonian quadratic in the phase space variables one can introduce in the classical system complex variables which are linear combinations of the phase space variables such that the hamiltonian can be written as a linear combination of the squared modulus of these complex variables. When the complex variables are replaced by deformed ladder operators one finds a deformed hamiltonian with a deformed spectrum and a deformed Fock space of eigenstates. The deformation of the Heisenberg algebra and the uncertainty principle can be interpreted as the replacement of Planck constant by an operator. Not all the symmetries of the classical system are compatible with the deformation. In some cases one can use the criteria to respect the symmetries of the classical theory in the classical-quantum transition to fix some ambiguities in the identification of the complex variables corresponding to the deformed ladder operators. 

\medskip

One could go beyond the systems with a quadratic hamiltonian considered in this work. Any hamiltonian which can be expanded in powers of the phase space variables can be reformulated at the classical level in terms of the complex variables defined by the quadratic part of the hamiltonian and then the replacement of the complex variables by the deformed ladder operators in the higher order terms defines the deformed quantum system. 

\medskip

Some extensions or applications of the proposal presented in this work can be considered. The main idea used to define a deformation of the classical-quantum transition in a quantum mechanical system, based on the introduction of a formulation of the classical system in terms of appropriately chosen complex variables which are replaced by deformed ladder operators in the quantum theory, can be easily extended to the case of field theory defining a deformed quantum field theory. Also the contraction of the discrete spectrum of the harmonic oscillator for large $n$ will have an analog for a many particle system with a large occupation number when the deformation of ladder operators is introduced in this context. One can guess that the deformation can be relevant in the determination of the critical temperature of  Bose-Einstein condensation.   

\medskip 

This work was supported by Spanish grants PGC2018-095328-B-I00 (FEDER/Agencia estatal de investigación), and DGIID-DGA No. 2015-E24/2 (J.L.C.) and Dicyt
041831GR (J. G.). One of us (J. G.) thanks the Alexander von Humboldt Foundation by the support.


%

\end{document}